\documentclass[12pt,english,preprint]{article}
\usepackage[T1]{fontenc}
\usepackage[latin9]{inputenc}
\usepackage{geometry}
\usepackage{amstext}
\usepackage{graphicx}
\usepackage{setspace}
\usepackage{esint}
\makeatletter
\newcommand{\lyxaddress}[1]{
\par {\raggedright #1
\vspace{1.4em}
\noindent\par}
}

\usepackage{slashed}

\makeatother

\usepackage{babel}

\begin{document}

\title{Large N Quantum Gravity\footnote{Contribution to appear in "New Journal of Physics Focus Issue on Quantum Einstein Gravity".}}

\author{Alessandro Codello%
\thanks{codello@sissa.it%
}}

\maketitle

\lyxaddress{\begin{center}
SISSA\\
via Bonomea 265, I-34136 Trieste, Italy
\par\end{center}}

\begin{abstract}
We obtain the effective action of four dimensional quantum gravity,
induced by $N$ massless matter fields, by integrating the RG flow
of the relative effective average action.
By considering the leading approximation in the large $N$ limit, where one neglects the gravitational contributions with respect
to the matter contributions, we show how different aspects
of quantum gravity, as asymptotic safety, quantum corrections to the
Newtonian potential and the conformal anomaly induced effective action, are all
represented by different terms of the effective action when this is
expanded in powers of the curvature.
\end{abstract}

\section{Introduction}

Even if we are still lacking a quantum theory of gravitational phenomena
we are starting to accumulate interesting partial results which will
probably be important bits of the final theory. These are, among others,
the conformal anomaly induced effective action, which was
first written down in \cite{Riegert_1984}; the low energy
corrections to the Newtonian gravitational potential analyzed in \cite{Donoghue_1994};
the possible ultraviolet (UV) finite completion of the theory described
in the asymptotic safety scenario \cite{Percacci_2009}.
In this paper we want to show how these different aspects can all be seen as arising
from different terms in the gravitational effective action when this is expanded
in powers of the curvature. In particular, the UV properties of the
theory are related to the zero and first order terms, i.e. to the
renormalization of Newton's and cosmological constants, and to the
renormalization of the coupling constants of all higher order local
invariants; quantum gravitational corrections to the Newtonian potential
are encoded in the finite part of the curvature square terms, while
the conformal anomaly induced effective action is just one of the possible
four curvature structures.

We will use the effective average action formalism to obtain
the gravitational effective action induced by 
$N_{0}$ massless scalar fields, $N_{\frac{1}{2}}$ massless Dirac spinors and $N_{1}$ abelian gauge fields,
interacting solely with the background geometry, as the result of the integration of the RG flow.
In the large $N$ expansion\footnote{$N$ is any of $N_{0}, N_{\frac{1}{2}}, N_{1}$} one assumes that the number
of matter fields $N$ grows large. In the leading approximation one simply neglects the gravitational contributions with respect
to the matter contributions: this eliminates computational and conceptual issues related to the treatment
of gravitational fluctuations. Within the effective average
action formalism this point of view has been analyzed in \cite{Percacci_2006}, where the focus was
on the local part of the effective average action and on the related
renormalizability issues. Here we will extend the large $N$ analysis
to a non-local truncation of effective average action where this is expanded in powers of the curvature.

\section{Effective average action for matter fields on curved space}

Following \cite{Percacci_2006} we consider massless matter fields on a curved four dimensional
manifold equipped with a metric $g_{\mu\nu}$. The bare action we consider describes $N_{0}$ massless scalar fields, $N_{\frac{1}{2}}$ massless Dirac spinors and $N_{1}$ 
abelian gauge fields interacting only with the background geometry:
\begin{eqnarray}
S[\phi,\psi,A_{\mu},\bar{c},c;g]&=&\int d^{4}x\sqrt{g}\left\{ \sum_{i=1}^{N_{0}}\left[\frac{1}{2}\partial_{\mu}\phi_{i}\partial^{\mu}\phi_{i}+\frac{\chi}{12}\phi_{i}^{2}R\right]+\sum_{i=1}^{N_{\frac{1}{2}}}\bar{\psi}_{i}\slashed\nabla\psi_{i}\qquad\right.\nonumber\\
&&\qquad\qquad\left.+\sum_{i=1}^{N_{1}}\left[\frac{1}{4}F_{\mu\nu,i}F_{i}^{\mu\nu}+\frac{1}{2\alpha}\left(\partial_{\mu}A_{i}^{\mu}\right)^{2}+\partial_{\mu}\bar{c}_{i}\,\partial^{\mu}c_{i}\right]\right\} \,.\label{1}
\end{eqnarray}
Here $\chi$ is a parameter, only when $\chi=1$ the scalar action is conformally invariant if the scalar
has conformal weight one. The Dirac operator is defined using the covariant Dirac matrices $\gamma^{\mu}=e_{a}^{\mu}\gamma^{a}$.
Note also that on an arbitrary curved manifold the abelian ghosts do not
decouple and cannot be discarded; we will choose the gauge $\alpha=1$ from
now on.

The effective average action $\Gamma_{k}[\phi,\psi,A_{\mu},\bar{c},c;g]$
is a scale depended generalization of the standard effective action
(depending on the infrared (IR) cutoff scale $k$) that interpolates
smoothly between the bare action for $k\rightarrow\infty$ and the
full quantum effective action for $k\rightarrow0$ \cite{Wetterich_1993}.
It satisfies an exact RG flow equation
describing its flow as the IR scale $k$ is shifted. The flow equation
relevant to the field content we are considering reads as follows:
\begin{equation}
\partial_{t}\Gamma_{k}[\varphi;g]=\frac{1}{2}\mathrm{Tr}\frac{\partial_{t}R_{k}[g]}{\Gamma_{k}^{(2,0)}[\varphi;g]+R_{k}[g]}\,,\label{1.2}
\end{equation}
where $\Gamma_{k}^{(2,0)}[\varphi,g]$ is the Hessian of the effective
average action taken with respect to the field multiplet $\varphi=(\phi,\psi,A_{\mu},\bar{c},c)$ and the trace is a "super-trace" on this space.
Equation (\ref{1.2}) is exact and is both UV and IR finite. The flow
equation (\ref{1.2}) can be seen as a RG improvement of the one-loop
effective action which is derived from the modified bare action $S[\varphi,g]\rightarrow S[\varphi,g]+\Delta S_{k}[\varphi,g]$,
where $\Delta S_{k}[\varphi,g]$ is a cutoff action quadratic in the
fields. This cutoff action is constructed in such a way to suppress the propagation
of all field modes smaller than the RG scale $k$. The effective average
action and the exact RG flow it satisfies (\ref{1.2}) offer a different
approach to quantization: in theory space, the space of "all"
action functionals, the bare action represents the UV starting point
of a RG trajectory which reaches the quantum effective action in the
IR. The integration of successive modes is done step by step as the
cutoff scale $k$ is lowered. More details on the construction of
the effective average action on curved backgrounds or in presence
of quantized gravity can be found in \cite{Codello_Percacci_Rahmede_2009}.

As explained in \cite{Codello_2010}, since we are considering matter fields interacting only with the background geometry, we can replace the the Hessian of the effective average action in the rhs of (\ref{1.2}) with the Hessian of the bare action (\ref{1}). After performing the field multiplet trace we find\footnote{We are using a Type II cutoff operator in the nomenclature of \cite{Codello_Percacci_Rahmede_2009}.}:
\begin{eqnarray}
\partial_{t}\Gamma_{k}[\phi,\psi,A_{\mu},\bar{c},c;g] & = & \frac{N_{0}}{2}\textrm{Tr}_{0}\, h_{k}\left(\Delta_{0}\right)-\frac{N_{\frac{1}{2}}}{2}\textrm{Tr}_{\frac{1}{2}}\, h_{k}(\Delta_{\frac{1}{2}})\nonumber \\
 &  & +N_{1}\left[\frac{1}{2}\textrm{Tr}_{1}\, h_{k}\left(\Delta_{1}\right)-\textrm{Tr}_{0}\, h_{k}\left(\Delta_{gh}\right)\right]\,,\label{1.3}
\end{eqnarray}
where we defined the function $h_{k}(z)=\frac{\partial_{t}R_{k}(z)}{z+R_{k}(z)}$
and $\textrm{Tr}_{s}$ indicates a trace over spin-$s$ fields. The
differential operators introduced in (\ref{1.3}) are the following:
\begin{equation}
\Delta_{0}=\Delta+\frac{\chi}{6}R\quad\quad\quad\Delta_{\frac{1}{2}}=\Delta+\frac{1}{4}R\quad\quad\quad(\Delta_{1})^{\mu\nu}=\Delta g^{\mu\nu}+R^{\mu\nu}\quad\quad\quad\Delta_{gh}=\Delta\,,\label{1.4}
\end{equation}
where $\Delta=-g^{\mu\nu}\nabla_{\mu}\nabla_{\nu}$ is the covariant Laplacian. In the next section we will determine the RG flow of the effective average action by calculating the functional traces on the rhs of (\ref{1.3}).

\section{Flow equations and beta functions}

The traces on the rhs of the flow equation (\ref{1.3}) can be expanded in powers of the curvature by employing the the non-local
heat kernel expansion \cite{Barvinsky_Vilkovinsky_1990} as was done in \cite{Codello_2010,Satz_Codello_Mazzitelli_2010}.
To second order we find the following result%
\footnote{Here we define $\Gamma_{k}[g]\equiv\Gamma_{k}[0,0,0,0,0;g]$%
}:
\begin{eqnarray}
(4\pi)^{2}\,\partial_{t}\Gamma_{k}[g] & = & \frac{N_{0}-4N_{\frac{1}{2}}+2N_{1}}{2}Q_{2}[h_{k}]\int d^{4}x\sqrt{g}\nonumber \\
 &  & +\frac{(1-\chi)N_{0}+2N_{\frac{1}{2}}-4N_{1}}{12}Q_{1}[h_{k}]\int d^{4}x\sqrt{g}R\nonumber \\
 &  & +\int d^{4}x\sqrt{g}\, R\left[\int_{0}^{\infty}ds\,\tilde{h}_{k}(s)\, s\, F_{R}(s\Delta)\right]R\nonumber \\
 &  & +\int d^{4}x\sqrt{g}\, C_{\mu\nu\alpha\beta}\left[\int_{0}^{\infty}ds\,\tilde{h}_{k}(s)\, s\, F_{C}(s\Delta)\right]C^{\mu\nu\alpha\beta}\nonumber \\
 &  & -\frac{N_{0}+11N_{\frac{1}{2}}+62N_{1}}{720}Q_{0}[h_{k}]\int d^{d}x\sqrt{g}E\nonumber \\
 &  & +\frac{N_{0}+N_{\frac{1}{2}}-3N_{1}}{60}Q_{0}[h_{k}]\int d^{d}x\sqrt{g}\Delta R\nonumber \\
 &  & +O(\mathcal{R}^{3})\,,\label{2}
\end{eqnarray}
where $\mathcal{R}$ stands for any curvature. The non-local heat kernel
structure functions $F_{C}(x)$ and $F_{R}(x)$ in (\ref{2}) are
linear combinations of the basic non-local heat kernel structure function
$f(x)=\int_{0}^{1}d\xi\, e^{-x\xi(1-\xi)}$ and read as follows:
\begin{eqnarray}
F_{R}(x) & = & \frac{5N_{0}-20N_{\frac{1}{2}}+10N_{1}}{48}\frac{f(x)-1}{x^{2}}+\frac{(3-2\chi)N_{0}-2N_{\frac{1}{2}}-2N_{1}}{48}\frac{f(x)}{x}\nonumber \\
 &  & -\frac{(13-12\chi)N_{0}+8N_{\frac{1}{2}}-22N_{1}}{288}\frac{1}{x}+\frac{(3-2\chi)^{2}N_{0}-2N_{1}}{576}f(x)\nonumber \\
F_{C}(x) & = & \frac{N_{0}-4N_{\frac{1}{2}}+2N_{1}}{4}\frac{f(x)-1}{x^{2}}-\frac{N_{\frac{1}{2}}-2N_{1}}{4}\frac{f(x)}{x}\nonumber \\
 &  & +\frac{N_{0}+2N_{\frac{1}{2}}-10N_{1}}{24}\frac{1}{x}+\frac{N_{1}}{8}f(x)\,.\label{3}
 \end{eqnarray}
In  (\ref{2}) $\tilde{h}_{k}(s)$ is the (inverse) Laplace transform of the
function $h_{k}(x)$ and the $Q$-functionals are defined as $Q_{n}[f]=\frac{1}{\Gamma(n)}\int_{0}^{\infty}dz\, z^{n-1}f(z)$
when $n>0$ and as $Q_{n}[f]=(-1)f^{(n)}(0)$ when $n\leq0$.

Since the RG flow has generated all possible terms compatible with diffeomorphism invariance on the rhs of (\ref{2}),
we need now to consider the most general truncation for the gravitational effective average action $\Gamma_{k}[g]$ to insert in the lhs of (\ref{2}).
As proposed in \cite{Codello_2010,Satz_Codello_Mazzitelli_2010},
we consider an ansatz where the effective average action is expanded
in powers of the curvature and where the scale dependence is carried
both by running couplings and by running (possibly non-local) structure
functions. To second order in the curvatures the expansion of the gravitational
effective average reads as follows\footnote{This implyes also that we are adding to the action (\ref{1}) the
non-dynamical term $\Gamma_{\Lambda}[g]$.}:
\begin{eqnarray}
\Gamma_{k}[g] & = & \int d^{4}x\sqrt{g}\left[\frac{1}{16\pi G_{k}}(2\Lambda_{k}-R)+R\, f_{R,k}(\Delta)R+\right.\nonumber \\
 &  & \left.+C_{\mu\nu\alpha\beta}f_{C,k}(\Delta)C^{\mu\nu\alpha\beta}+\frac{1}{\rho_{k}}E+\frac{1}{\tau_{k}}\Delta R\right]+O(\mathcal{R}^{3})\,.\label{4}
\end{eqnarray}
In (\ref{4}) the couplings $\Lambda_{k}$ and $G_{k}$ are the running
cosmological and Newton's constants, $f_{C,k}$ and $f_{R,k}$
are the two independent curvature square running structure functions
(which at this point are arbitrary functions of the covariant Laplacian $\Delta$),
while $\rho_{k}$ and $\tau_{k}$ are the running couplings related
to the Euler and total derivative invariants. The other two curvature square running couplings $\lambda_{k}$ and $\xi_{k}$ are related
to the running structure functions by the following relations:
\begin{equation}
f_{C,k}(0)=\frac{1}{2\lambda_{k}}\qquad\qquad\qquad f_{R,k}(0)=\frac{1}{\xi_{k}}\,.\label{4.1}
\end{equation}

We can extract the beta functions for the couplings $\Lambda_{k},G_{k},\rho_{k},\tau_{k}$
and the flow equations for the running structure functions $f_{C,k}$,
$f_{R,k}$ by comparing (\ref{2}) with (\ref{4}). In particular,
by matching the coefficients of the operators $\int\sqrt{g}$ and
$\int\sqrt{g}R$ we find the following relations:
\begin{eqnarray}
(4\pi)^{2}\partial_{t}\left(\frac{\Lambda_{k}}{8\pi G_{k}}\right) & = & \frac{N_{0}-4N_{\frac{1}{2}}+2N_{1}}{2}Q_{2}[h_{k}]\nonumber \\
(4\pi)^{2}\partial_{t}\left(-\frac{1}{16\pi G_{k}}\right) & = & \frac{(1-\chi)N_{0}+2N_{\frac{1}{2}}-4N_{1}}{12}Q_{1}[h_{k}]\,,\label{5}
\end{eqnarray}
while by matching the non-local curvature square terms we find the following
flow equation for both running structure functions:
\begin{equation}
\partial_{t}f_{i,k}(x)=\frac{1}{(4\pi)^{2}}\int_{0}^{\infty}ds\,\tilde{h}_{k}(s)\, s\, F_{i}(sx)\,,\label{6}
\end{equation}
where $i=C,R$ and $x$ stands for $\Delta$. If we insert the
explicit form of the functions $F_{i}(x)$ from (\ref{3}) we can
rewrite equation (\ref{6}) explicitly in terms of $Q$-functionals:
\begin{eqnarray}
(4\pi)^{2}\,\partial_{t}f_{R,k}(x) & = & \frac{(3-2\chi)^{2}N_{0}-2N_{1}}{576}\int_{0}^{1}d\xi\, Q_{0}\left[h_{k}\left(z+x\xi(1-\xi)\right)\right]+\nonumber \\
 &  & +\frac{(3-2\chi)N_{0}-2N_{\frac{1}{2}}-2N_{1}}{48}\int_{0}^{1}d\xi\, Q_{1}\left[h_{k}\left(z+x\xi(1-\xi)\right)\right]\nonumber \\
 &  & -\frac{(13-12\chi)N_{0}+8N_{\frac{1}{2}}-22N_{1}}{288}Q_{1}[h_{k}]\nonumber \\
 &  & +\frac{5N_{0}-20N_{\frac{1}{2}}+10N_{1}}{48x^{2}}\left\{ \int_{0}^{1}d\xi\, Q_{2}\left[h_{k}\left(z+x\xi(1-\xi)\right)\right]-Q_{2}[h_{k}]\right\} \quad\label{10}
\end{eqnarray}
\begin{eqnarray}
(4\pi)^{2}\,\partial_{t}f_{C,k}(x) & = & \frac{N_{1}}{8}\int_{0}^{1}d\xi\, Q_{0}\left[h_{k}\left(z+x\xi(1-\xi)\right)\right]+\nonumber \\
 &  & -\frac{N_{\frac{1}{2}}-2N_{1}}{4x}\int_{0}^{1}d\xi\, Q_{1}\left[h_{k}\left(z+x\xi(1-\xi)\right)\right]+\nonumber \\
 &  & +\frac{N_{0}+2N_{\frac{1}{2}}-10N_{1}}{24x}Q_{1}[h_{k}]+\nonumber \\
 &  & +\frac{N_{0}-4N_{\frac{1}{2}}+2N_{1}}{4x^{2}}\left\{ \int_{0}^{1}d\xi\, Q_{2}\left[h_{k}\left(z+x\xi(1-\xi)\right)\right]-Q_{2}[h_{k}]\right\} \,.\quad\label{11}\end{eqnarray}
Finally, matching the remaining local curvature square terms gives:
\begin{eqnarray}
(4\pi)^{2}\partial_{t}\left(\frac{1}{\rho_{k}}\right) & = & -\frac{N_{0}+11N_{\frac{1}{2}}+62N_{1}}{360}Q_{0}[h_{k}]\nonumber \\
(4\pi)^{2}\partial_{t}\left(\frac{1}{\tau_{k}}\right) & = & \frac{N_{0}+N_{\frac{1}{2}}-3N_{1}}{30}Q_{0}[h_{k}]\,.\label{11.1}
\end{eqnarray}
The flow equations (\ref{5}), (\ref{10}), (\ref{11}) and (\ref{11.1})
completely describe the RG flow of the effective average action (\ref{4}).

The $Q$-functionals in the flow equations just derived can be evaluated
once a particular cutoff shape function has been chosen; it is possible
to evaluate them analytically if we employ the so called "optimized" cutoff shape
function $R_{k}(z)=(k^{2}-z)\theta(k^{2}-z)$. We easily find the
cutoff shape independent value $Q_{0}\left[h_{k}\right]=2$ and the
results $Q_{n}\left[h_{k}\right]=\frac{2}{n!}k^{2n}$, while the evaluation
of the parametric integrals of $Q$-functional present in (\ref{10})
and (\ref{11}) is more involved and we refer to the Appendix of \cite{Satz_Codello_Mazzitelli_2010}
for further details. 
We can now write down the beta functions for the dimensionless cosmological
constant, $\Lambda_{k}=k^{2}\tilde{\Lambda}_{k}$, and for the dimensionless
Newton's constant, $G_{k}=k^{2-d}\tilde{G}_{k}$, which follow from
(\ref{5}):
\begin{eqnarray}
\partial_{t}\tilde{\Lambda}_{k} & = & -2\tilde{\Lambda}_{k}+\frac{N_{0}-4N_{\frac{1}{2}}+2N_{1}}{4\pi}\tilde{G}_{k}+\frac{(1-\chi)N_{0}+2N_{\frac{1}{2}}-4N_{1}}{6\pi}\tilde{\Lambda}_{k}\tilde{G}_{k}\nonumber \\
\partial_{t}\tilde{G}_{k} & = & 2\tilde{G}_{k}+\frac{(1-\chi)N_{0}+2N_{\frac{1}{2}}-4N_{1}}{6\pi}\tilde{G}_{k}^{2}\,.\label{11.2}
\end{eqnarray}
The first terms on the rhs of (\ref{11.2}) represent the canonical scaling of the cosmological and Newton's constants,
while the other terms represent the running induced by the interaction with matter fields.
Note also that conformal scalars, i.e. $\chi=1$, do not contribute to the running of Newton's constant.
Equations (\ref{10}) and (\ref{11}) can be compactly rewritten as:
\begin{equation}
\partial_{t}f_{i,k}(x)=\frac{1}{(4\pi)^{2}}g_{i}\left(\frac{x}{k^{2}}\right)\,,\label{12}
\end{equation}
where we defined the following cutoff shape dependent functions:
\begin{equation}
g_{i}(u)\equiv\int_{0}^{\infty}ds\,\tilde{h}_{k}(s)\, s\, F_{i}(sk^{2}u)\,.\label{12.1}
\end{equation}
In (\ref{12.1}) $u=x/k^{2}$ stands for the covariant Laplacian in units of $k$; note that the UV regime corresponds to small values of $u$ while the IR regime corresponds to large values of $u$.
The integrals in (\ref{12.1}) can be evaluated analytically \cite{Satz_Codello_Mazzitelli_2010},
we find the following forms:
\begin{eqnarray}
g_{C}(u) & = & \frac{N_{0}+6N_{\frac{1}{2}}+12N_{1}}{120}-\left[\frac{N_{0}+6N_{\frac{1}{2}}+12N_{1}}{120}-\frac{8N_{0}+8N_{\frac{1}{2}}-64N_{1}}{120u}\right.\nonumber \\
 &  & \qquad\qquad\left.+\frac{16N_{0}-64N_{\frac{1}{2}}+32N_{1}}{120u^{2}}\right]\sqrt{1-\frac{4}{u}}\theta(u-4)\nonumber \\
g_{R}(u) & = & \frac{(1-\chi)^{2}N_{0}}{72}-\left[\frac{(1-\chi)^{2}N_{0}}{72}+\frac{(1-\chi)N_{0}+N_{\frac{1}{2}}-2N_{1}}{18u}\right.\nonumber \\
 &  & \qquad\qquad\left.+\frac{N_{0}-4N_{\frac{1}{2}}+2N_{1}}{18u^{2}}\right]\sqrt{1-\frac{4}{u}}\theta(u-4)\,.\label{13}
\end{eqnarray}
Note that the form of (\ref{13}) implies that conformally invariant
matter, contributes to $g_{R}(u)$ and thus
to the flow of $f_{R,k}(x)$, showing that for $k\neq0$ the flow generates
non-conformally invariant interactions. We will see later that the conformal
invariance of the curvature square terms will be partially restored in the IR at $k=0$. The
constant terms in (\ref{13}), when matched with (\ref{4.1}), give
the beta functions for the couplings $\lambda_{k}$ and $\xi_{k}$.
These, together with the explicit forms of the beta functions (\ref{11.1}),
are:
\begin{eqnarray}
\partial_{t}\lambda_{k} & = & -\frac{1}{(4\pi)^{2}}\frac{N_{0}+6N_{\frac{1}{2}}+12N_{1}}{60}\lambda_{k}^{2}\nonumber \\
\partial_{t}\xi_{k} & = & -\frac{1}{(4\pi)^{2}}\frac{(1-\chi)^{2}N_{0}}{72}\xi_{k}^{2}\nonumber \\
\partial_{t}\rho_{k} & = & \frac{1}{(4\pi)^{2}}\frac{N_{0}+11N_{\frac{1}{2}}+62N_{1}}{180}\rho_{k}^{2}\nonumber \\
\partial_{t}\tau_{k} & = & -\frac{1}{(4\pi)^{2}}\frac{N_{0}+N_{\frac{1}{2}}-3N_{1}}{15}\tau_{k}^{2}\label{15}
\end{eqnarray}
The numerical coefficients in (\ref{15}) are scheme independent, i.e. they don't depend on $R_{k}(z)$ and on the other details of the cutoff choice.
Note that the beta function of the coupling $\xi_{k}$ vanishes when one is considering
conformally invariant matter. The beta functions (\ref{11.2}) and (\ref{15})
agree with those obtained in \cite{Percacci_2006}.

\section{Asymptotic safety}

We proceed now to integrate the RG flow of the couplings of the local
curvature invariants present in the truncation we are considering,
i.e. the cosmological, Newton's, and the coupling constants of the curvature
square terms. We will see that precisely the flow of these couplings
is related to the UV renormalization of the theory, that in the effective
average action formalism is encoded into the initial conditions of
RG flow. These relations will tell us how to choose the bare couplings
in order to achieve a finite continuum limit when the UV regularization is removed. 

Since (\ref{11.2}) is a closed system for $\tilde{\Lambda}_{k}$
and $\tilde{G}_{k}$, we can start to look for non-Gaussian fixed
points of the RG flow, that we need to find in order to construct a continuum
limit, by first solving:
\begin{equation}
\partial_{t}\tilde{\Lambda}_{k}=0\qquad\qquad\partial_{t}\tilde{G}_{k}=0\,.\label{17}
\end{equation}
It's easy to see that the system (\ref{17}) admits both a Gaussian
fixed point $\tilde{\Lambda}_{k}=\tilde{G}_{k}=0$ and a non-Gaussian
one \cite{Percacci_2006}:
\begin{figure}
\centering{}\includegraphics[scale=0.7]{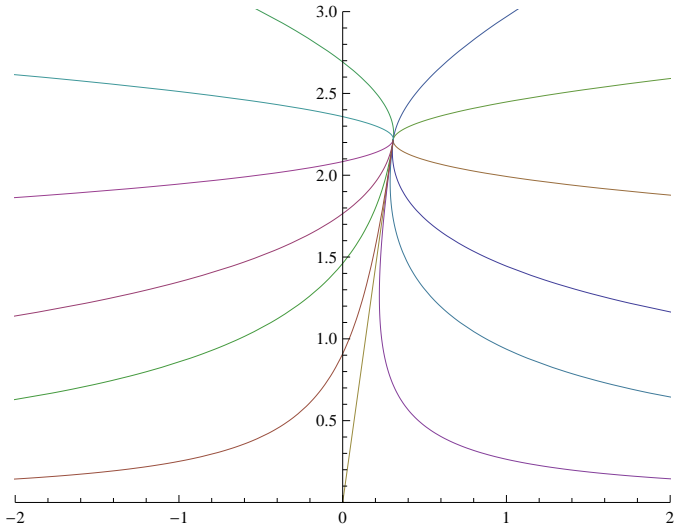}
\caption{Renormalization group flow in the $(\tilde{\Lambda}_{k},\tilde{G}_{k})$
plane form equations (\ref{18.4}) and (\ref{18.71}).}
\end{figure}
\begin{equation}
\tilde{\Lambda}_{*}=-\frac{3}{4}\frac{N_{0}-4N_{\frac{1}{2}}+2N_{1}}{(1-\chi)N_{0}+2N_{\frac{1}{2}}-4N_{1}}\qquad\qquad\tilde{G}_{*}=-\frac{12\pi}{(1-\chi)N_{0}+2N_{\frac{1}{2}}-4N_{1}}\,.\label{18}
\end{equation}
Note that to have a fixed point with positive Newton's constant we
need to satisfy the inequality $(1-\chi)N_{0}+2N_{\frac{1}{2}}-4N_{1}>0$. The stability
matrix around the non-Gaussian fixed point (\ref{18}) is readily calculated
and shows that it is attractive in the UV \footnote{It has eigenvalues $-4,-2$.}. This shows that both the cosmological and Newton's constants are
asymptotically safe couplings. This fact shows that, within the the truncation we are considering and in the large $N$ limit,
four dimensional quantum gravity is asymptotically safe\footnote{See \cite{Codello_Percacci_Rahmede_2009} for a discussion of asymptotic
safety when also gravitational fluctuations are considered.}. The beta function system (\ref{11.2}) can be easily integrated analytically.
Integrating the second equation in (\ref{11.2}) from the UV scale
$\Lambda$ to the IR scale $k$ gives the following relation connecting
the $k$-dependent, fixed point and bare Newton's constants%
\footnote{If we define the couplings $A_{k}=\frac{\Lambda_{k}}{8\pi G_{k}}$ and
$B_{k}=\frac{1}{16\pi G_{k}}$ the system (\ref{11.2}) becomes $\partial_{t}\tilde{A}_{k}=4(\tilde{A}_{*}-\tilde{A}_{k})$
and $\partial_{t}\tilde{B}_{k}=2(\tilde{B}_{*}-\tilde{B}_{k})$ where
$A_{k}=\tilde{A}_{k}k^{4}$ and $B_{k}=\tilde{B}_{k}k^{2}$. These relations
immediately give $\tilde{A}_{k}=\tilde{A}_{*}+(\tilde{A}_{\Lambda}-\tilde{A}_{*})\left(\frac{\Lambda}{k}\right)^{4}$
and $\tilde{B}_{k}=\tilde{B}_{*}+(\tilde{B}_{\Lambda}-\tilde{B}_{*})\left(\frac{\Lambda}{k}\right)^{2}$
from which we read off (\ref{18.1}) and (\ref{18.6}).}:
\begin{equation}
\frac{1}{\tilde{G}_{k}}=\frac{1}{\tilde{G}_{*}}+\left(\frac{1}{\tilde{G}_{\Lambda}}-\frac{1}{\tilde{G}_{*}}\right)\left(\frac{\Lambda}{k_{0}}\right)^{2}\left(\frac{k_{0}}{k}\right)^{2}\,.\label{18.1}
\end{equation}
In (\ref{18.1}) we introduced the reference scale $k_{0}$, which
will play a role similar to the renormalization scale $\mu$ of standard
perturbation theory. To make (\ref{18.1}) finite in the continuum limit, i.e. in the limit $\Lambda\rightarrow\infty$,
we need to renormalize Newton's constant; this can be done by
imposing the following condition:
\begin{equation}
\left(\frac{1}{\tilde{G}_{\Lambda}}-\frac{1}{\tilde{G}_{*}}\right)\left(\frac{\Lambda}{k_{0}}\right)^{2}=C_{G}\,,\label{18.2}
\end{equation}
where $C_{G}$ is a constant that we will determine later. Solving
(\ref{18.2}) with respect to bare Newton's constant $\tilde{G}_{\Lambda}$
gives:
\begin{equation}
\tilde{G}_{\Lambda}=\frac{\tilde{G}_{*}}{1+C_{G}\tilde{G}_{*}\left(\frac{k_{0}}{\Lambda}\right)^{2}}\,,\label{18.3}\end{equation}
which shows how we need to choose $\tilde{G}_{\Lambda}$ to approach
$\tilde{G}_{*}$, as $\Lambda\rightarrow\infty$, in order that the
lhs of (\ref{18.2}) remains constant. Inserting now (\ref{18.2})
in (\ref{18.1}) gives the functional form of the $k$-dependent Newton's
constant:\begin{equation}
\tilde{G}_{k}=\frac{\tilde{G}_{*}}{1+C_{G}\tilde{G}_{*}\left(\frac{k_{0}}{k}\right)^{2}}\,.\label{18.4}\end{equation}
Equations (\ref{18.3}) and (\ref{18.4}) are formally the same at
the level of the approximation we are considering but have different
physical meanings: the first equation tells us how the bare coupling changes
as we vary the UV scale $\Lambda$, while the second equation shows how the
$k$-dependent coupling flows as we integrate more and more degrees
of freedom by lowering the RG scale $k$. From (\ref{18.4}) we see
that the dimensionful Newton's constant, $G_{k}=k^{-2}\tilde{G}_{k}$,
reaches the following renormalized value at $k=0$:
\begin{equation}
G_{0}=\frac{1}{C_{G}k_{0}^{2}}\qquad\Rightarrow\qquad C_{G}=\frac{1}{G_{0}k_{0}^{2}}\,,\label{18.5}
\end{equation}
which fixes the value of the constant introduced (\ref{18.2}).
Similarly, we obtain the following relation between the $k$-dependent,
fixed point and bare cosmological constants:
\begin{equation}
\frac{\tilde{\Lambda}_{k}}{\tilde{G}_{k}}=\frac{\tilde{\Lambda}_{*}}{\tilde{G}_{*}}+\left(\frac{\tilde{\Lambda}_{\Lambda}}{\tilde{G}_{\Lambda}}-\frac{\tilde{\Lambda}_{*}}{\tilde{G}_{*}}\right)\left(\frac{\Lambda}{k_{0}}\right)^{4}\left(\frac{k_{0}}{k}\right)^{4}\,.\label{18.6}
\end{equation}
After we impose the UV renormalization condition:\[
\left(\frac{\tilde{\Lambda}_{\Lambda}}{\tilde{G}_{\Lambda}}-\frac{\tilde{\Lambda}_{*}}{\tilde{G}_{*}}\right)\left(\frac{\Lambda}{k_{0}}\right)^{4}=C_{\Lambda}\,,\]
which can be solved for the bare cosmological constant:
\begin{equation}
\tilde{\Lambda}_{\Lambda}=\frac{\tilde{\Lambda}_{*}+C_{\Lambda}\tilde{G}_{*}\left(\frac{k_{0}}{\Lambda}\right)^{4}}{1+C_{G}\tilde{G}_{*}\left(\frac{k_{0}}{\Lambda}\right)^{2}}\,,\label{18.7}
\end{equation}
we obtain the $k$-dependent cosmological constant:
\begin{equation}
\tilde{\Lambda}_{k}=\frac{\tilde{\Lambda}_{*}+C_{\Lambda}\tilde{G}_{*}\left(\frac{k_{0}}{k}\right)^{4}}{1+C_{G}\tilde{G}_{*}\left(\frac{k_{0}}{k}\right)^{2}}\,.\label{18.71}
\end{equation}
As before, (\ref{18.7}) and (\ref{18.71}) are formally equivalent.
The renormalized value of the dimensionful cosmological constant $\Lambda_k=k^{2}\tilde{\Lambda}_k$ is obtained from (\ref{18.71}) and shows that $C_{\Lambda}=\frac{\Lambda_{0}}{G_{0}k_{0}^{4}}$.
Since both $G_{k}$ and $\Lambda_{k}$ are relevant parameters in
the UV (they are both attracted by the non-Gaussian fixed point)
we need to fix their value from experiments, i.e. we need to measure
$\Lambda_{0}/k_{0}^{2}$ and $G_{0}k_{0}^{2}$. The phase diagram
obtained by plotting equations (\ref{18.4}) and (\ref{18.71}) is
shown in Fig. 1; we can clearly see that all flow trajectories reach the
IR without any obstruction, a feature that is not yet present in the analogous
RG flows that include also gravitational fluctuations \cite{Codello_Percacci_Rahmede_2009}.
Note that the dimensionless product $\Lambda_{0}G_{0}$ is independent
of the arbitrary scale $k_{0}$, thus $\tilde{\Lambda}_{0}\tilde{G}_{0}$
is a true IR numerical prediction associated to a given RG trajectory.
The Einstein-Hilbert terms in our ansatz (\ref{4}), in the $k\rightarrow0$
limit becomes:
\begin{equation}
\left.\Gamma_{0}[g]\right|_{\mathcal{R}^{0}+\mathcal{R}^{1}}=\frac{\Lambda_{0}}{8\pi G_{0}}\int d^{4}x\sqrt{g}-\frac{1}{16\pi G_{0}}\int d^{4}x\sqrt{g}R\,.\label{19.3}
\end{equation}
Obviously this part of the gravitational effective action is not conformally
invariant due to the presence of the scales $\Lambda_{0}$ and $G_{0}$.
We see from (\ref{18.1}) and (\ref{18.6}) that the only way to obtain
conformal invariance in the IR is to set the UV couplings to their
fixed point values: $\tilde{\Lambda}_{*}=\tilde{\Lambda}_{\Lambda}$
and $\tilde{G}_{\Lambda}=\tilde{G}_{*}$ so that $\Lambda_{0}=0$
and $1/G_{0}=0$, obviously the only way to restore conformal
invariance at the quantum level. 

Next we can integrate the beta functions (\ref{15}) for the couplings $\lambda_{k},\xi_{k},\rho_{k},\tau_{k}$. These couplings
have only the Gaussian fixed point, at least within the approximation
we are considering \cite{Benedetti_Machado_Saueressig_2009}. This
is easily done:
\[\frac{1}{g_{k}}=\frac{1}{g_{\Lambda}}+C\,\log\frac{\Lambda}{k}=\frac{1}{g_{\Lambda}}+C\,\log\frac{\Lambda}{k_{0}}-C\,\log\frac{k}{k_{0}}\,,\]
where $g_{k}$ stands for any of the couplings $\lambda_{k},\xi_{k},\rho_{k},\tau_{k}$
and where the relative constants $C$ can be read-off from (\ref{15}).
We can now choose the following UV boundary conditions:
\begin{equation}
g_{\Lambda}=-\frac{1}{C\,\log\frac{\Lambda}{k_{0}}}\,,\label{20}
\end{equation}
which imply the following $k$-dependence of the coupling constants:
\begin{equation}
g_{k}=-\frac{1}{C\,\log\frac{k}{k_{0}}}\,.\label{21}
\end{equation}
As we mentioned before, (\ref{20}) and (\ref{21}) are formally equivalent; they both show that these couplings are asymptotically free.
Note that when we reinsert the $g_{k}$ from (\ref{21}) in the effective average
action (\ref{4}) and we take the limit $k\rightarrow0$ the local
curvature square contributions diverge. It seems that the effective
action has an IR divergence. We will see in the next section that
this is not the case: when we include the finite curvature square
contributions in (\ref{4}), the $\log k$ in (\ref{21}) combines with another
such term, coming from the running structure functions, so that their sum has a finite $k\rightarrow0$ limit.

\section{Curvature square terms and corrections to the Newtonian potential}

In this section we start to look at the finite non-local part of the
effective action and we show how this part of the action can be obtained as the result
of the integration of the RG flow. We focus on the running structure functions in the curvature square
terms, where we can perform all steps analytically. 

By integrating the flow equation (\ref{12}) for the running structure
functions from the UV scale $\Lambda$ to the IR scale $k$ we find
the relation:
\[f_{i,\Lambda}(x)-f_{i,k}(x)=\frac{1}{(4\pi)^{2}}\int_{k}^{\Lambda}\frac{dk'}{k'}g_{i}\left(\frac{x}{k'^{2}}\right)\,,\]
where $f_{i,\Lambda}(x)$ are the bare structure functions, while $f_{i,k}(x)$ are the $k$-dependent ones. Changing variables to $u=x/k^{2}$, with $dk/k=-du/2u$,
gives:
\begin{equation}
f_{i,k}(x)=f_{i,\Lambda}(x)-\frac{1}{(4\pi)^{2}}\int_{x/\Lambda^{2}}^{x/k^{2}}\frac{du}{2u}g_{i}\left(u\right)\,.\label{25}
\end{equation}
It should be noted that when we insert the explicit forms of the functions $g_{i}(u)$
from (\ref{13}), the constant terms will make the integrals in (\ref{25})
logarithmically divergent at the lower limit when $\Lambda\rightarrow\infty$.
We can isolate these divergencies by subtracting these constants from
the integrals in (\ref{25}) in the following way:
\[f_{C,k}(x)=f_{C,\Lambda}(x)-\frac{1}{(4\pi)^{2}}\frac{N_{0}+6N_{\frac{1}{2}}+12N_{1}}{120}\log\frac{\Lambda}{k_{0}}\]
\begin{equation}
+\frac{1}{(4\pi)^{2}}\frac{N_{0}+6N_{\frac{1}{2}}+12N_{1}}{120}\log\frac{k}{k_{0}}-\frac{1}{(4\pi)^{2}}\int_{x/\Lambda^{2}}^{x/k^{2}}\frac{du}{2u}\left[g_{C}\left(u\right)-\frac{N_{0}+6N_{\frac{1}{2}}+12N_{1}}{120}\right]\label{25.1}
\end{equation}
and
\[f_{R,k}(x)=f_{R,\Lambda}(x)-\frac{1}{(4\pi)^{2}}\frac{(1-\chi)^{2}N_{0}}{72}\log\frac{\Lambda}{k_{0}}\]
\begin{equation}
+\frac{1}{(4\pi)^{2}}\frac{(1-\chi)^{2}N_{0}}{72}\log\frac{k}{k_{0}}-\frac{1}{(4\pi)^{2}}\int_{x/\Lambda^{2}}^{x/k^{2}}\frac{du}{2u}\left[g_{R}\left(u\right)-\frac{(1-\chi)^{2}N_{0}}{72}\right]\,.\label{26}
\end{equation}
The $\log\frac{\Lambda}{k_{0}}$ terms in the first lines of (\ref{25.1})
and (\ref{26}) are the UV divergences that are also encountered in standard
perturbation theory \cite{Shapiro_2008}. We can renormalize the running structure functions by
imposing the following UV boundary conditions:
\begin{eqnarray}
f_{C,\Lambda}(x) & = & \frac{1}{(4\pi)^{2}}\frac{N_{0}+6N_{\frac{1}{2}}+12N_{1}}{120}\log\frac{\Lambda}{k_{0}}+c_{C}\nonumber \\
f_{R,\Lambda}(x) & = & \frac{1}{(4\pi)^{2}}\frac{(1-\chi)^{2}N_{0}}{72}\log\frac{\Lambda}{k_{0}}+c_{R}\,,\label{27}
\end{eqnarray}
that make the first lines in (\ref{25.1}) and (\ref{26}) vanish. Here $c_{C}$ and $c_{R}$ are possible finite renormalizations constants.
In this way the second lines of (\ref{25.1}) and (\ref{26}) are
finite in the UV $\Lambda\rightarrow\infty$ and in the IR $k\rightarrow0$
limits. Note also that the first terms of the second lines of (\ref{25.1})
and (\ref{26}), when combined with (\ref{4.1}) correctly reproduce
(\ref{21}). Performing the integrals in (\ref{25.1}) and (\ref{26})
now gives the following finite contributions to the curvature square part of the renormalized effective action:
\begin{eqnarray}
\left.\Gamma_{0}[g]\right|_{\mathcal{R}^{2}} & = & -\frac{1}{2(4\pi)^{2}}\int d^{4}x\sqrt{g}\left[\frac{N_{0}+6N_{\frac{1}{2}}+12N_{1}}{120}C_{\mu\nu\alpha\beta}\log\frac{\Delta}{k_{0}^{2}}C^{\mu\nu\alpha\beta}\right.\nonumber \\
 &  &\qquad\qquad \left.+\frac{(1-\chi)^{2}N_{0}}{72}R\,\log\frac{\Delta}{k_{0}^{2}}R\right]\,,\label{28}
 \end{eqnarray}
where we fixed the renormalization constants to the values:
\[c_{C}=-\frac{23N_{0}}{1800}-\frac{3N_{\frac{1}{2}}}{50}-\frac{4N_{1}}{75}\qquad\qquad c_{R}=\frac{1+10(1-\chi)+30(1-\chi)^{2}N_{0}}{2160}+\frac{N_{\frac{1}{2}}}{360}-\frac{N_{1}}{120}\,.\]
The effective action (\ref{28}) is the same as the one obtained using standard
perturbation theory if we equate the arbitrary scale $k_0$
with the renormalization scale $\mu$ \cite{Shapiro_2008}.
In (\ref{28}) we dropped the Euler and the total derivative terms,
as we mentioned in the previous section, these terms are not finite
in the IR limit $k\rightarrow0$ since their running is of the form
(\ref{21}), but this does not cause any problem since these terms
are total derivatives and thus vanish in general\footnote{The integral $\int\sqrt{g}E=32\pi^{2}\chi$ is the Euler characteristic
of the four dimensional manifold and is thus non-zero in general.
The renormalization of this term is probably related to topological
fluctuations.}. 
Note also that if we were employing a different cutoff shape function $R_{k}(z)$ than the one employed here, we would had found a different form for the flow for  $k\neq0$ but, as
was shown in the two dimensional case \cite{Codello_2010}, the
$k\rightarrow0$ limit would had  always been equal to (\ref{28}).

In (\ref{28}) the only contribution to the renormalized Ricci structure function $f_{R,0}(x)$
comes from the scalar fields and this contributions vanishes if we consider conformally invariant scalars by setting $\chi=1$.
Even if $\chi=1$ the functional (\ref{28}) is not conformally invariant. This happens because we are considering an ansatz of the effective average action
structured as in (\ref{4}), which by construction cannot accommodate
a conformally invariant limit for $k=0$ since the expansion we are performing is not so. As we will explain
in the following section, we expect that when $\chi=1$ the only non-conformally invariant part of
the effective action to be the conformal anomaly induced effective action \cite{Mazur_Mottola_2001}. The
form found in equation (\ref{28}) must thus be the result of the expansion
of a more general conformally invariant term, that to order curvature
square reduces to (\ref{28}). We can speculate such a term to be
of the following form:
\begin{equation}
\int d^{4}x\sqrt{g}\, C_{\mu\nu\alpha\beta}\,\log \mathcal{O}_{C}\,C^{\mu\nu\alpha\beta}\,,\label{29}
\end{equation}
where the operator $\mathcal{O}_{C}$ is a fourth order conformally invariant
differential operator that acts on four indices tensors with the symmetries
of the Weyl tensor. The explicit form of this differential operator is unknown \cite{Hamada_2001} but it must have a form such that (\ref{29}) has a smooth flat space limit.
At this point it is important to note that a conformal
variation of (\ref{28}), when $\chi=1$, will generate the anomaly term
proportional to $C^{2}$ with the correct coefficient (see equation (\ref{45})). This fact has
already been noticed by several authors \cite{Shapiro_2008, Deser_Schwimmer_1996_Deser_1996},
in particular the second authors consider this a genuine anomaly term
in contrast to the expectation that effective action proposed in  \cite{Riegert_1984} is the only
one that produces the conformal anomaly upon variation \cite{Mazur_Mottola_2001}.
Since these first authors obtained the form (\ref{28}) by a flat
space calculation, we can think of them to be in the same position as we are in our
calculation: the non-conformally invariant form, and thus anomalous
term, is what remains of a truly conformally invariant term of the form
(\ref{29}) that reduces to a non-invariant term when, to be able to perform calculations, an expansion
that brakes conformal symmetry is performed. This point of view can compose the
different expectation expressed on one side by \cite{Deser_Schwimmer_1996_Deser_1996}
and on the other side by \cite{Mazur_Mottola_2001} and thus deserves
further study.

A physical reason to expect a term like (\ref{28}) in the effective action, at second order
in the curvatures, is that it generates the quantum gravitational
corrections to Newton's gravitational potential\footnote{This is only the "graviton polarization" part of the correction.}. These corrections can be obtained
along the lines of \cite{Satz_Codello_Mazzitelli_2010} and are the
following:
\begin{equation}
V(r)=-\frac{MG_{0}}{r}\left\{ 1+\left(\frac{43}{30\pi}+\frac{\left[4+5(1-\chi)^{2}\right]N_{0}+24N_{\frac{1}{2}}+48N_{1}}{180\pi}\right)\frac{G_{0}}{r^{2}}\right\} \,.\label{30}
\end{equation}
The first part of the quantum gravitational correction in (\ref{30}) is the
purely gravitational one found in \cite{Satz_Codello_Mazzitelli_2010}.
As explained in this last reference,  the correction (\ref{30}) is equivalent to the one obtained using effective field theory arguments.
Note that every field species (even conformally invariant matter) gives a positive contribution, of the same form of the purely gravitational one. From (\ref{30}) we can see that when the number of massless matter fields $N$ grows large, their contributions soon dominate the purely gravitational contribution.

\section{Higher order terms and the conformally induced effective action}

In the last two sections we explicitly derived and integrated the RG flow of the gravitational effective average action to second order in the curvature expansion.
In this section we now look at the third and fourth order terms.

One can consider the curvature cube terms in the effective average action ansatz (\ref{4}) and can project their flow 
by employing in (\ref{2}) the non-local heat
kernel expansion to third order \cite{Barvinsky_Gusev_Vilkovisky_Zhytnikov_1994_2009}.
There are now ten independent running structure functions and the
curvature expansion of the effective average action has the following form:
\begin{equation}
\left.\Gamma_{k}[g]\right|_{\mathcal{R}^{3}}=\frac{1}{(4\pi)^{2}}\sum_{i=1}^{10}\int d^{d}x\sqrt{g}\, f_{i,k}(\Delta_{1},\Delta_{2},\Delta_{3})\mathcal{R}_{1}\mathcal{R}_{2}\mathcal{R}_{3}(i)\,,\label{40}
\end{equation}
where the curvature structures $\mathcal{R}_{1}\mathcal{R}_{2}\mathcal{R}_{3}(i)$
are given in \cite{Barvinsky_Gusev_Vilkovisky_Zhytnikov_1994_2009}.
For example $\mathcal{R}_{1}\mathcal{R}_{2}\mathcal{R}_{3}(1)=R_{1}R_{2}R_{3}$,
$\mathcal{R}_{1}\mathcal{R}_{2}\mathcal{R}_{3}(2)=R_{1\mu}^{\alpha}R_{2\nu}^{\mu}R_{3\alpha}^{\nu}$
and so on\footnote{Note that we are considering only the purely gravitational curvature
structures of \cite{Barvinsky_Gusev_Vilkovisky_Zhytnikov_1994_2009}.}; following the conventions of \cite{Barvinsky_Gusev_Vilkovisky_Zhytnikov_1994_2009} the Laplacians $\Delta_{i}$ act only on the curvatures $\mathcal{R}_{i}$
for $i=1,2,3$. The third order terms on the rhs of the flow equation
(\ref{2}) can now be written as:
\begin{equation}
\left.\partial_{t}\Gamma_{k}[g]\right|_{\mathcal{R}^{3}}=\frac{1}{(4\pi)^{2}}\sum_{i=1}^{10}\int_{0}^{\infty}ds\,\tilde{h}_{k}(s)\, s^{\alpha_{i}-1}\int d^{d}x\sqrt{g}\, F_{i}(s\Delta_{1},s\Delta_{2},s\Delta_{3})\mathcal{R}_{1}\mathcal{R}_{2}\mathcal{R}_{3}(i)\,,\label{41}
\end{equation}
where the explicit form of the non-local heat kernel structure functions
$F_{i}(x_{1},x_{2},x_{3})$ can be obtained from those given in \cite{Barvinsky_Gusev_Vilkovisky_Zhytnikov_1994_2009}.
Since the curvature structures $\mathcal{R}_{1}\mathcal{R}_{2}\mathcal{R}_{3}(i)$
may contain covariant derivatives, third order terms are proportional
to different powers of $s$. In fact we have $\alpha_{1}=\alpha_{2}=\alpha_{3}=3$,
$\alpha_{4}=\alpha_{5}=\alpha_{6}=\alpha_{7}=4$, $\alpha_{8}=\alpha_{9}=5$
and $\alpha_{10}=6$. By comparing (\ref{40}) to (\ref{41}) we
obtain the flow equations for the third order running structure functions:
\begin{equation}
\partial_{t}f_{i,k}(x_{1},x_{2},x_{3})=\frac{1}{(4\pi)^{2}}g_{i}\left(\frac{x_{1}}{k^{2}},\frac{x_{3}}{k^{2}},\frac{x_{3}}{k^{2}}\right)\,,\label{42}
\end{equation}
where we introduced the scheme dependent functions:
\begin{equation}
g_{i}(u_{1},u_{2},u_{3})=\frac{1}{(4\pi)^{2}}\int_{0}^{\infty}ds\,\tilde{h}_{k}(s)\, s^{\alpha_{i}-1}F_{i}(sk^{2}u_{1},sk^{2}u_{2},sk^{2}u_{3})\,,\label{43}
\end{equation}
where $u_{i}=x_{i}/k^{2}$ for $i=1,2,3$.
At this order, computations become increasingly hard and it seems
rather difficult to straightforwardly evaluate the $s$ integrals on
the rhs of (\ref{43}), even is we employ the optimized cutoff shape function $R_{k}(z)=(k^2-z)\theta(k^2-z)$.
It was shown in \cite{Percacci_2006} that if one employes this cutoff shape function, the couplings of the cubic and higher invariants,
i.e. $g_{i,k}\equiv g_{i,k}(0,0,0)$, have zero beta functions.
With this cutoff choice, we need to UV renormalize only the couplings we already considered at lower order of the curvature expansion.
In spite of this, the structure functions $g_{i}(x_1,x_2,x_3)$ are non-zero even if they have 
a vanishing Taylor expansion around the origin (since $Q_{-n}[h_{k}]=0$ for all $n\ge$1), probably they are proportional to theta functions as are the second order functions $g_{i}(u)$ (\ref{13}). 
For this reason no UV divergencies arise when we integrate the flow equation (\ref{43}) and we find a finite result for the renormalized third order structure functions:
\begin{equation}
f_{i,0}(x_{1},x_{2},x_{3})=\frac{1}{(4\pi)^{2}}\int_{0}^{\infty}\frac{dk}{k}g_{i}\left(\frac{x_{1}}{k^{2}},\frac{x_{3}}{k^{2}},\frac{x_{3}}{k^{2}}\right)\,.\label{44}
\end{equation}
In (\ref{44}) we imposed the UV boundary conditions $f_{i,\Lambda}(x_{1},x_{2},x_{3})=0$.
The absence of UV divergencies can also be understood by looking at the
local expansion of the gravitational effective average action where, as we just noticed, the dimensionful beta functions
turn out to be zero $\partial_{t}g_{i,k}=0$, thus implying $\tilde{g}_{i,k}=\left(\frac{\Lambda}{k}\right)^{d_{i}}\tilde{g}_{i,\Lambda}$
where $g_{i,k}=\tilde{g}_{i,k}$ are the dimensionless coupling with $d_{i}<0$. Thus in the limit $\Lambda\rightarrow\infty$ is finite and we have $g_{i,k}\rightarrow0$.
We finally obtain the cubic curvature part of the gravitational effective
action $\left.\Gamma_{0}[g]\right|_{\mathcal{R}^{3}}$ when we re-insert
(\ref{44}) into (\ref{40}).

It will be interesting to extend the curvature expansion to the next (quartic)
order in the curvatures, since the conformal anomaly induced effective action is of this order \cite{Riegert_1984}.
This effective action is usually obtained by integrating the conformal anomaly;
for the field content we are considering and for $\chi=1$ this is the following:
\begin{eqnarray}
\left\langle T_{\mu}^{\mu}\right\rangle  & = & \frac{1}{(4\pi)^{2}}\left[\frac{N_{0}+6N_{\frac{1}{2}}+12N_{1}}{120}C^{2}
-\frac{N_{0}+11N_{\frac{1}{2}}+62N_{1}}{360}E\right.\nonumber \\
 &  & \left.\qquad\qquad+\frac{N_{0}+N_{\frac{1}{2}}-3N_{1}}{60}\Delta R\right]\,.\label{45}
\end{eqnarray}
To derive (\ref{45}) one uses the relation $\left\langle T_{\mu}^{\mu}\right\rangle =(4\pi)^{-2}\int\sqrt{g}\textrm{tr}\, b_{4}(S^{(2)})$,
where $S^{(2)}$ is the Hessian of (\ref{1}) and the $b_{4}$ are
the relative local heat kernel coefficients. It was shown in \cite{Riegert_1984}
that the following effective action has an energy-momentum tensor whose trace reproduces (\ref{45}):
\begin{eqnarray}
\Gamma_{CA}[g] & = & \frac{1}{8(4\pi)^{2}}\int d^{4}x\sqrt{g}\left[\frac{N_{0}+6N_{\frac{1}{2}}+12N_{1}}{60}C^{2}\right.\nonumber \\
&  & \left.-\frac{N_{0}+11N_{\frac{1}{2}}+62N_{1}}{360}\left(E+\frac{2}{3}\Delta R\right)\right]\frac{1}{\Delta_{4,0}}\left(E+\frac{2}{3}\Delta R\right)\,,\label{46}
\end{eqnarray}
where $\Delta_{4,0}=\Delta^{2}+2R^{\mu\nu}\nabla_{\mu}\nabla_{\nu}+\frac{2}{3}R\Delta+\frac{1}{3}\nabla^{\mu}R\nabla_{\mu}$
is the conformal covariant fourth-order operator acting on conformal invariant scalars.
The conformal anomaly induced effective action (\ref{45}) will constitute, together with other conformally invariant terms, the four curvature terms in the curvature expansion of the effective action. 
But since to quartic order in the curvatures we don't even know the form of the non-local heat kernel
expansion, any calculation along the lines of the previous sections is, for the moment, a very hard task to perform.
In spite of this we believe that the integration of the flow of the effective average action will generate
the non-local effective action (\ref{45}) as one of the many fourth order terms.
In  \cite{Codello_2010} it was shown that this is indeed what happens in the analogous two dimensional case. 

\section{Conclusions}

As a first step toward understanding the gravitational effective
action we studied the effective action induced by massless
matter fields interacting solely with the background geometry.
We proposed an expansion of the effective action
in powers of the curvatures and we showed how various known results
about quantum gravity could be seen as arising from different terms
of it. We showed that asymptotic safety and UV renormalization are
related to the running of the couplings of local curvature
invariants, the first of these being the cosmological and Newton's
constants. Then we derived the quantum gravitational corrections to the Newtonian potential
from the quadratic curvature terms and we explained that 
all other quantum gravitational effects, including those induced by
the conformal anomaly are contained in the higher order contributions.

Once matter contributions are well understood, one can infer something
about the gravitational effective action when matter
contributions become dominant in the large $N$ limit.
When instead gravitational fluctuations become large, we must consider the fact that
gravity may actively reacts to matter inducing corrections to the
effective action that may be of  the same order as those induced by matter.
The exact solution of two dimensional quantum gravity shows that this 
is indeed what happens in the two dimensional
case \cite{Ambjorn:1997di}. It is thus important to extend this work
and the program started in \cite{Satz_Codello_Mazzitelli_2010} to the treatment of
full gravitational dynamics.

\section*{Acknowledgments}

I would like to thank R. Percacci for careful reading the manuscript
and for useful and stimulating discussions.


\begin{thebibliography}{22}

\bibitem{Riegert_1984}R.J. Riegert, Phys. Lett. B $\mathbf{134}$ (1984) 56;\\
E.S. Fradkin, A.A. Tseytlin, Phys. Lett. B $\mathbf{134}$ (1984)  187.

\bibitem{Donoghue_1994}J.F. Donoghue, Phys. Rev. D 50 (1994) 3874,
\mbox{arXiv:gr-qc/9405057}.

\bibitem{Percacci_2009}R. Percacci in \textit{Approaches to Quantum
Gravity}, D. Oriti (Ed.), Cambridge University Press (2009), \mbox{arXiv:0709.3851 [hep-th]}.

\bibitem{Wetterich_1993}C. Wetterich, Phys. Lett. B $\mathbf{301}$
(1993) 90; J.~Berges, N. Tetradis and C. Wetterich, Phys. Rep. $\mathbf{363}$
(2002) 223.

\bibitem{Codello_Percacci_Rahmede_2009}A. Codello, R. Percacci and
C. Rahmede, Annals Phys. $\mathbf{324}$ (2009) 414, arXiv:0805.2909
{[}hep-th{]}.

\bibitem{Percacci_2006}R. Percacci, Phys. Rev. D $\mathbf{73}$ (2006)
041501, arXiv:hep-th/0511177.


\bibitem{Barvinsky_Vilkovinsky_1990}A. O. Barvinsky and G. A. Vilkovisky,
Nucl. Phys. B $\mathbf{333}$ (1990) 471; I.G. Avramidi, Nucl. Phys. B $\mathbf{355}$
(1991) 712. Erratum-ibid. B $\mathbf{509}$ (1998) 557{]}; G.A. Vilkovisky, {}``Heat kernel: Rencontre
entre physiciens et mathematiciens'', CERN-TH-6392-92.

\bibitem{Codello_2010}A. Codello, Annals Phys. 325 (2010) 1727, \mbox{arXiv:1004.2171 [hep-th]}.

\bibitem{Satz_Codello_Mazzitelli_2010}A. Satz, A. Codello and F.D.
Mazzitelli, Phys. Rev. D 82 (2010) 084011, \mbox{arXiv:1006.3808 [hep-th]}.

\bibitem{Benedetti_Machado_Saueressig_2009} D. Benedetti, P. Machado
and F. Saueressig, Mod. Phys. Lett. A $\mathbf{24}$ (2009) 2233;
Nucl.Phys. B $\mathbf{824}$ (2010) 168. 

\bibitem{Shapiro_2008}I. L. Shapiro, Class. Quant. Grav. $\mathbf{25}$
(2008) 103001, arXiv:0801.0216 {[}gr-qc{]}.

\bibitem{Hamada_2001}K.j. Hamada, Prog. Theor. Phys. $\mathbf{105}$
(2001) 673, arXiv:hep-th/0012053.

\bibitem{Deser_Schwimmer_1996_Deser_1996}S. Deser and A. Schwimmer,
Phys. Lett. B $\mathbf{309}$ (1993) 279, arXiv:hep-th/9302047; S.
Deser, Helv. Phys. Acta $\mathbf{69}$ (1996) 570, arXiv:hep-th/9609138.




\bibitem{Mazur_Mottola_2001}P.O. Mazur and E. Mottola, Phys. Rev.
D $\mathbf{64}$ (2001) 104022, arXiv:hep-th/0106151.

\bibitem{Barvinsky_Gusev_Vilkovisky_Zhytnikov_1994_2009}A.O. Barvinsky,
Yu.V. Gusev, G.A. Vilkovisky and V.V. Zhytnikov, J. Math. Phys. $\mathbf{35}$
(1994) 3525, arXiv:gr-qc/9404061; A.O. Barvinsky, Yu.V. Gusev, V.V.
Zhytnikov and G.A. Vilkovisky, arXiv:0911.1168 {[}hep-th{]}.

\bibitem{Ambjorn:1997di}
  J. Ambjorn, B. Durhuus and T. Jonsson,
  {\it Quantum geometry. A statistical field theory approach},
Cam. Univ. Press (1997).

\end{thebibliography}
\end{document}